\documentclass[%
 reprint,
 amsmath,amssymb,
 aps,
]{revtex4-1}

\usepackage[dvips]{color}
\usepackage{graphicx}
\usepackage{dcolumn}
\usepackage{bm}

\newcommand{\nc}{\newcommand}
\nc{\tcr}{\textcolor{red}}

\begin{document}

\preprint{APS/123-QED}

\title{Exact Relaxation Dynamics in the
Totally Asymmetric Simple Exclusion Process}

\author{Kohei Motegi$^1$}
\author{Kazumitsu Sakai$^2$}
\author{Jun Sato$^3$}
\affiliation{%
$^1$Okayama Institute for Quantum Physics, 
 Kyoyama 1-9-1, Okayama 700-0015, Japan \\
$^2$Institute of physics, University of Tokyo, 
Komaba 3-8-1, Meguro-ku, Tokyo 153-8902, Japan \\
$^3$Department of Physics, Graduate School of Humanities and Sciences,
Ochanomizu University 2-1-1 Ohtsuka, Bunkyo-ku, Tokyo 112-8610, Japan
}%

\date{\today}

\begin{abstract}
The relaxation dynamics of the one-dimensional totally asymmetric simple 
exclusion process on a ring is considered in the case of step initial 
condition. Analyzing the time evolution of the local particle densities
and currents by the Bethe ansatz method, we examine their full relaxation
dynamics. As a result, 
we observe peculiar behaviors, such as the emergence of a ripple in the density profile 
and the existence of the excessive particle currents.
Moreover, by making a finite-size scaling analysis of
the asymptotic amplitudes of 
the local densities and currents,
we find the scaling exponents with respect to the total number of
sites to be $-3/2$ and $-1$ respectively.
\begin{description}
\item[PACS numbers]
05.60.Cd, 02.50.Ey, 75.10.Pq
\end{description}
\end{abstract}

\pacs{Valid PACS appear here}
\maketitle

{\it Introduction.}$-$
Nonequilibrium statistical mechanics has been a hot topic for the last twenty
years. Although a large amount of research has been done, the theory beyond 
the linear response regime is still far from being complete. One strategy to gain 
insights into nonequilibrium statistical mechanics is to examine special 
models in which exact methods can be employed, and to extract as much 
information as possible from them.

Among them, the asymmetric simple exclusion process
(ASEP) \cite{De,Sch,Spit,Li} is one of the well-studied paradigms in 
nonequilibrium statistical mechanics. The ASEP is a continuous time 
Markov process describing the diffusion of classical particles on a 
one-dimensional lattice. Each particle is viewed as a biased random 
walker that obeys an exclusion principle: each lattice site can not 
be occupied by more than one particle. 
This simple model appears in many different contexts. It originally appeared 
to describe the dynamics of ribosomes along RNA \cite{MGP},
and have modern applications to pedestrian and traffic flows 
\cite{Scha,SCN}, transport of quantum dots \cite{KO}.
It can also be regarded as a discrete analogue of the KPZ equation
\cite{KPZ}, which describes the surface growth phenomena,
observed in recent experiments on electroconvection \cite{TS}.

The ASEP is rather a toy, but  plenty of exact results have been obtained 
and still continuously provides valuable insights into nonequilibrium 
statistical mechanics. For instance, as for the steady state properties,
the matrix product ansatz was used to reveal various interesting phenomena 
such as the boundary induced phase transitions \cite{DEHP,Sasa,BECE,Kr}. 
On the other hand, as for the dynamical properties, the asymptotic behavior
of the relaxation process was exactly investigated by the Bethe ansatz 
\cite{Dh,GS,K,GM,deGE0,deGE1,AKSS,deGE2},
and shown that it is governed by the KPZ 
universality class.  In addition, the current fluctuation, which is
known to be described by the largest eigenvalue distribution of random matrices,
is  recently studied by use of random matrix theory and 
the Bethe ansatz
 \cite{Jo,PS,RS,IS,TW,DL,deGE2}. Despite these remarkable developments,
it is still a challenging problem to examine the exact relaxation dynamics
beyond the first principles Monte Carlo simulations which requires many samples 
to obtain reliable results.

In this article, we study the exact dynamics of the totally asymmetric simple 
exclusion process (TASEP) on a periodic ring, which is a special case of the ASEP
where particles are allowed to move only in one direction. Starting from the step 
initial condition where the half of the system is consecutively occupied by the 
particles and the other half is empty,  we investigate the relaxation process to 
the steady state by use of the Bethe ansatz method.
By quantitively evaluating local particle densities and currents,
we observe interesting behaviors peculiar to the step initial condition
such as the emergence of a ripple in the density profile
and excessive flow of currents.
This is the first study to evaluate the full relaxation dynamics of the ASEP 
by using the Bethe ansatz method.
Moreover, by making a finite-size scaling analysis of
the asymptotic amplitudes of 
the local densities and currents,
we find the scaling exponents with respect to the total number of
sites to be $-3/2$ and $-1$ respectively.
The results are consequences of the advantages of the Bethe ansatz method
which allows us to conduct the finite-size scaling of the amplitudes
completely separate from the exponential
parts determining the relaxation times.

\begin{figure}[t]
\begin{center}
\includegraphics[width=0.3\textwidth]{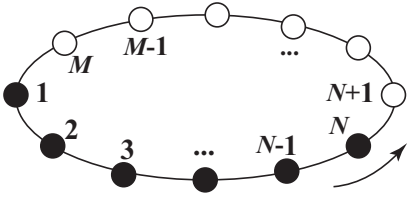}
\end{center}
\caption{The step initial condition of the TASEP.
TASEP is a special case of the ASEP where
particles are allowed to move only in one direction.}
\label{initialcondition}
\end{figure}

{\it Definition of the TASEP.$-$}
We consider the TASEP on a periodic ring with $M$ sites and $N$ particles.
Each site can be occupied by at most one particle (see Fig.~\ref{initialcondition}).
The dynamical rules of the TASEP is as follows: during the time interval $d t$, 
a particle at a site $j$ jumps to 
$j+1$th site with probability $dt$, if the $j+1$th site is 
empty.  A Boolean variable $\tau_i$ is associated to 
every site $j$ to indicate whether a particle is present ($\tau_i=1$) or 
not ($\tau_i=0$). The probability of being in the (normalized) state 
$|\tau_1, \dots, \tau_M \rangle$
is denoted as $P_t(\tau_1, \dots, \tau_M)$.
The time evolution of the state vector
$|\psi(t) \rangle=\sum_{\tau_i=0,1} P_t(\tau_1, \dots, \tau_M)
| \tau_1, \dots, \tau_M \rangle$
is subject to the master equation
\begin{align}
\frac{d}{d t} |\psi(t) \rangle
=\mathcal{M} |\psi(t) \rangle.
\label{master}
\end{align}
Here the Markov matrix $\mathcal{M}$ of the TASEP 
is defined by
\begin{align}
\mathcal{M}=\sum_{j=1}^M \Bigg\{
\sigma_j^+ \sigma_{j+1}^-
+\frac{1}{4}(\sigma_j^z \sigma_{j+1}^{z}-1)
 \Bigg\}, \label{markov}
\end{align}
where $\sigma_j^\pm:=(\sigma_j^x\pm i\sigma_j^y)/2$ and
$\sigma_j^{x,y,z}$ is the Pauli matrices acting on the $j$th site.
(Note here that we interpret that $\sigma_j^z |\tau_1,\dots, \tau_M\rangle=(-1)^{\tau_j} 
|\tau_1,\dots, \tau_M\rangle$).
The dynamical rules of the TASEP are encoded in this Markov matrix.
To describe the relaxation dynamics, we must evaluate the eigenvalues 
as well as the eigenstates of the Markov matrix $\mathcal{M}$.
The algebraic Bethe ansatz is one of the most useful methods
to achieve this.

{\it Algebraic Bethe Ansatz for Dynamics}.$-$
The Markov matrix $\mathcal{M}$ \eqref{markov} can be interpreted
as a (non-Hermitian) Hamiltonian of a one-dimensional quantum spin 
system which is known to be integrable. Therefore various exact methods
developed in the study of  integrable systems are also
available to analyze the dynamics of the TASEP.

To describe the relaxation dynamics by the Master equation
\eqref{master}, one must calculate both the eigenvalues and eigenstates
of \eqref{master} because  the time evolution of  the physical quantities 
are, in general, given by the form factors of   quantum operators
(see \eqref{expectation}).
To this end, we employ the algebraic Bethe ansatz \cite{KBI,Bo} which makes
it possible to calculate the correlation functions and  form factors of 
quantum integrable systems.

What plays a fundamental role is the
$L$-operator acting on the $j$th site
\begin{align}
L(j|u)
=us s_j+n(uI-u^{-1} s_j)+\sigma^- \sigma_j^+
+\sigma^+ \sigma_j^-,
\label{loperator}
\end{align}
where $s_j=(1+\sigma_j^z)/2$,
$n_j=(1-\sigma_j^z)/2$ is the projection operator
onto the empty and filled states at $j$th site,
respectively.
The monodromy matrix is defined by a product of $L$-operators:
\begin{align}
T(u)=\prod_{j=1}^M L(j|u)
&=
\begin{pmatrix}
A(u) & B(u)  \\
C(u) & D(u)
\end{pmatrix}.
\end{align}
The arbitrary $N$-particle state $|\psi \rangle$ (resp. its dual $\langle \psi|$) 
(not normalized)
is constructed by a multiple action
of $B$ (resp. $C$) operator on the vacuum state $|\Omega \rangle:=|0,\dots, 0\rangle$
(resp. $\langle \Omega|:=\langle0,\dots, 0|$):
\begin{align}
|\psi \rangle=\prod_{i=1}^N B(u_i)| \Omega \rangle,\quad
\langle \psi|=\langle \Omega| \prod_{i=1}^N
C(u_i).
\end{align} 
It can be shown that the above states are eigenstates of the Markov matrix 
\eqref{markov},  if the spectral parameters $\{ u \}=\{u_1, u_2, \dots, u_N \}$
satisfy the Bethe ansatz equation
\begin{align}
(1-u_k^{-2})^{-M} u_k^{-2N}=(-1)^{N-1}
\prod_{j=1}^N u_j^{-2}
\label{BAE}
\end{align}
for $k=1,2, \cdots,N$.
Then the eigenvalues of the 
Markov matrix  are given by
\begin{align}
\mathcal{M}=\sum_{j=1}^N \frac{1}{u_j^2-1}.
\end{align}

Now we formulate the relaxation dynamics of the TASEP within the algebraic 
Bethe ansatz. The time evolution of the expectation 
value for the physical quantity $\mathcal{A}$ starting from an initial 
state $| I_N \rangle$ is defined as
\begin{align}
\langle \mathcal{A} \rangle_t=\langle S_N| \mathcal{A} 
e^{\mathcal{M}t} | I_N \rangle.
\label{expectation}
\end{align}
Here $| S_N \rangle$ (resp. $\langle S_N|$) is the $N$-particle 
steady state (resp. its dual) which is simply constructed by 
the superposition of all configurations with equal probabilities.
We shall study the relaxation dynamics starting from
the step initial condition (see Fig.~\ref{initialcondition}) 
 where the half of the system is consecutively occupied by the 
particles and the other half is empty. The (normalized) initial state
$|I_N\rangle
=|\underbrace{1,\dots,1}_N,\underbrace{0,\dots,0}_{M-N}\rangle$
is given by $|I_N\rangle=B(1)^N|\Omega \rangle$
(see \cite{MC} for the XXZ spin chain).
(Note that this initial state is {\em not} the eigenstate
of the Markov matrix).
Thus the local densities $\langle n_i \rangle_t=\langle 1-s_i \rangle_t$
and currents $\langle j_i \rangle_t=\langle (1-s_i)s_{i+1} \rangle_t$
are respectively given by
\begin{align}
&\langle n_i \rangle_t
=\frac{N}{M}
+\sum_{\alpha}
\frac{e^{\mathcal{M}_\alpha t} (\langle S_N| \psi_\alpha \rangle-\langle S_N| s_i | 
\psi_\alpha \rangle)\langle \psi_\alpha|I_N \rangle}
{\langle \psi_\alpha|\psi_\alpha \rangle},
 \nonumber \\
&\langle j_i \rangle_t
=\frac{N(M-N)}{M(M-1)}  
 \label{currentexpansion}
\\
&\quad+\sum_{\alpha}
\frac{e^{\mathcal{M}_\alpha t}(\langle S_N| s_{i+1} | \psi_\alpha \rangle
-\langle S_N| s_i s_{i+1} | \psi_\alpha \rangle)
\langle \psi_\alpha|I_N \rangle} 
{\langle \psi_\alpha|\psi_\alpha \rangle}. \nonumber
\end{align}
Here the sum in the above is performed over all states except for
the steady state (denoted by the index $\alpha$).

The remaining problem is the evaluation of the norms and the scalar products of 
the Bethe vector, and the form factors of the local operators.
These quantities can be calculated in the framework of the algebraic Bethe ansatz
(cf. \cite{Bo}). For the norm of the Bethe vector $\langle \psi |\psi\rangle$,
one obtains
\begin{equation}
\langle \psi|\psi \rangle 
=\prod_{j=1}^N u_j^{2(N+M-1)}
\prod_{\substack{l,n=1 \\ l \neq n}}^N \frac{1}{u_l^2-u_n^2}
\mathrm{det}_N Q, \label{norm} 
\end{equation}
with $N\times N$ matrix 
\begin{equation}
Q_{jl}=\frac{N-1+(M-N+1) u_j^{-2}}{1-u_j^{-2}} \delta_{jl}
-(1-\delta_{jl}).  
\end{equation}
On the other hand, the overlap between the initial state and
the arbitrary Bethe vector is reduced to the following simple form: 
\begin{align}
&\langle \psi |I_N\rangle
=\prod_{j=1}^N (u_j-u^{-1}_j)^{M-N}u_j^{N-1}. \label{formone}
\end{align}
Finally the form factor for the local operators is explicitly given by
\begin{align}
	&\langle S_N|s_i \cdots s_{i+k-1}|
 \psi \rangle
=\prod_{j=1}^N(1-u_j^{-2})^{k+i-1} \prod_{j=1}^N u_j^{M+1}
\nonumber \\
&\times \prod_{N \ge l > n \ge 1}
\frac{1}{u_l^2-u_n^2} \mathrm{det}_N V^{(M-k)}, \label{formtwo} 
\end{align}
where the $N\times N$ matrix $V$ is written as
\begin{align}
V_{jl}^{(M-k)}=&\sum_{n=0}^{j-1} (-1)^n \frac{(M-k)!}{n!(M-k-n)!} 
u_l^{2(j-1-n)}, \nonumber
\end{align}
for $1 \le j \le N-1$ and
\begin{align}
V_{Nl}^{(M-k)}=&-\sum_{n=N-1}^{M-k} (-1)^n \frac{(M-k)!}{n!(M-k-n)!}
u_l^{-2(n-N+1)}. \nonumber
\end{align}
Note that the overlap between the steady state and the
Bethe vector 
$\langle S_N| \psi \rangle$
is obtained by setting
 $i=1, k=0$ in \eqref{formtwo}.
Inserting all of the above formula into \eqref{currentexpansion},
one can calculate  the time evolution of the expectation values of 
the local particle densities and currents.

\begin{figure}[tt]
\begin{center}
\includegraphics[width=0.4\textwidth]{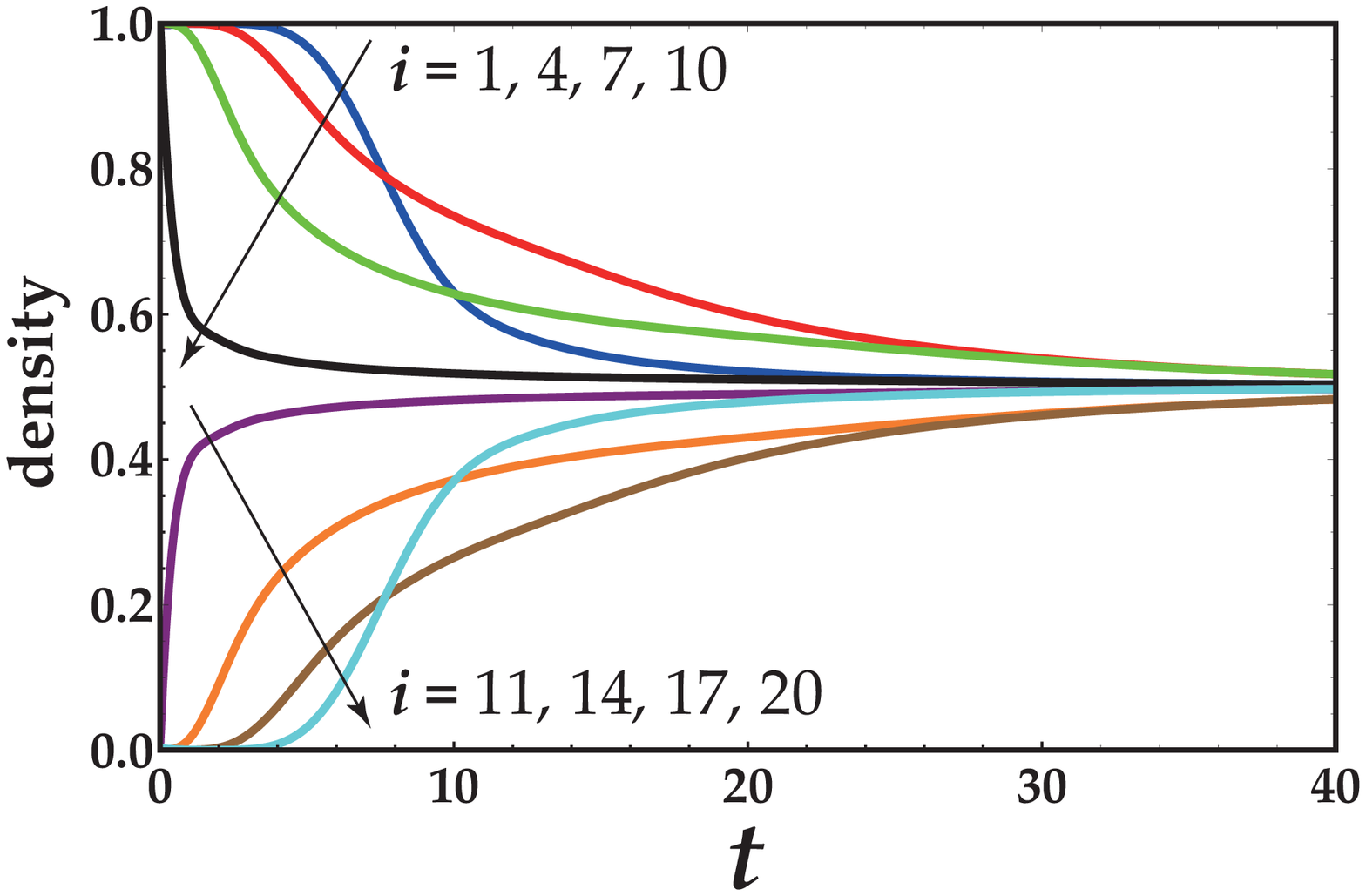}
\includegraphics[width=0.4\textwidth]{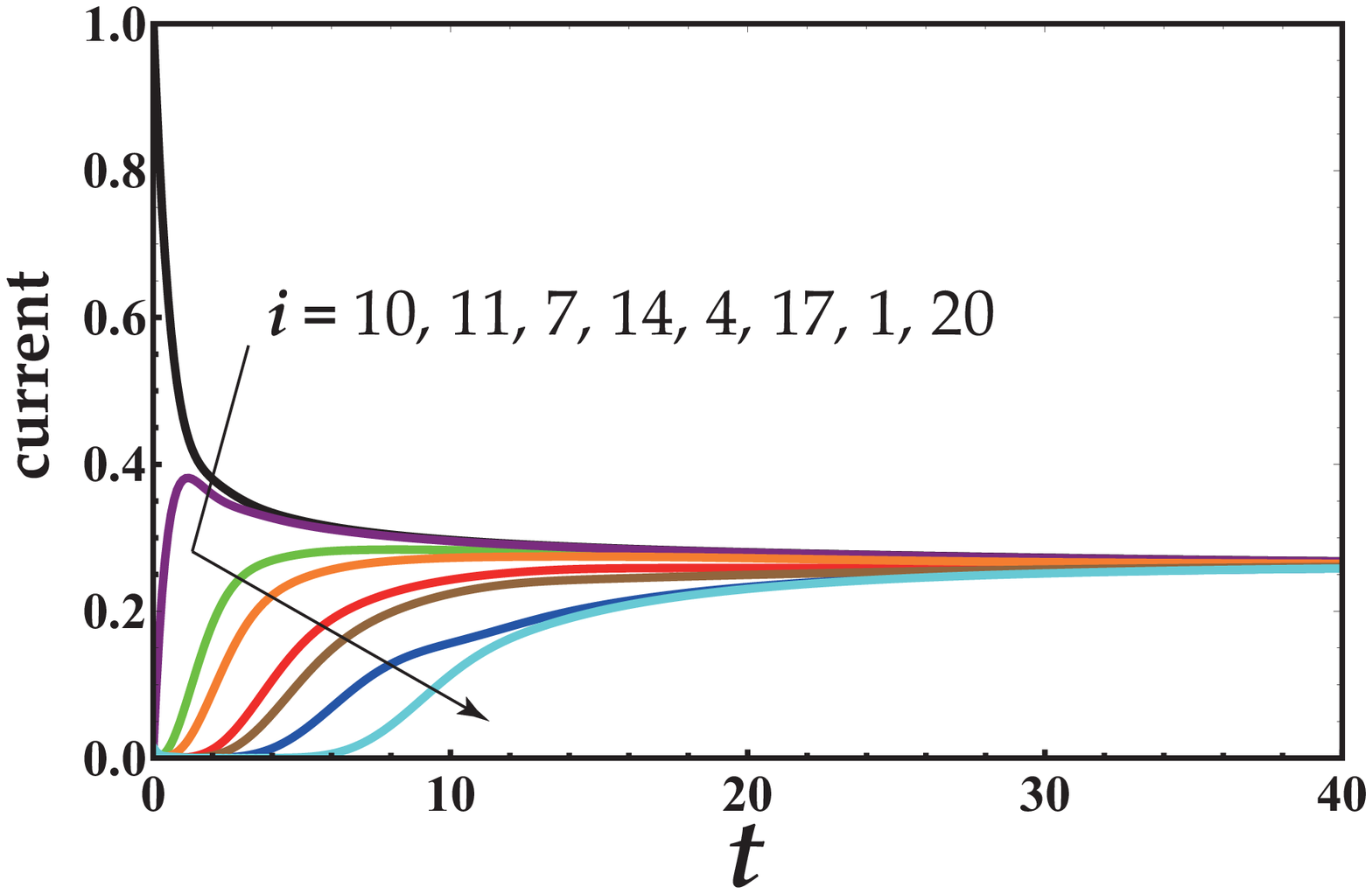}
\end{center}
\caption{(Color) The time evolution of local particle densities
$\langle n_i \rangle_t$ and currents
$\langle j_i \rangle_t$ for 20 sites, 10 particles.}
\label{timeevolution}
\end{figure}

{\it Results.}$-$
We use the algorithm in \cite{GM} to compute the
roots of the Bethe ansatz equation
\eqref{BAE} for the arbitrary states as many as possible,
and insert into the formulas
 \eqref{currentexpansion},
\eqref{norm}, \eqref{formone} and \eqref{formtwo}
to evaluate the local densities and currents.
We shall give a remark here for the algorithm in \cite{GM},
which conjectures that one choice of a monotonous function
gives one Bethe roots.
We find this one-to-one correspondence does not hold
anymore for $M \ge 10$ at half-filling.
There may be two or more or zero Bethe roots for one choice
of quantum numbers.
However, these irregular Bethe roots 
give highly excited states.
The missing roots are about 1\% and
does not not seem to contribute much to the
physical quantities, as can be seen from
the fact that the density profile of the
initial state is almost completely realized.
Here we demonstrate the case for 20 sites and 10 particles
(half-filling).

\begin{figure}[tt]
\begin{center}
\includegraphics[width=0.4\textwidth]{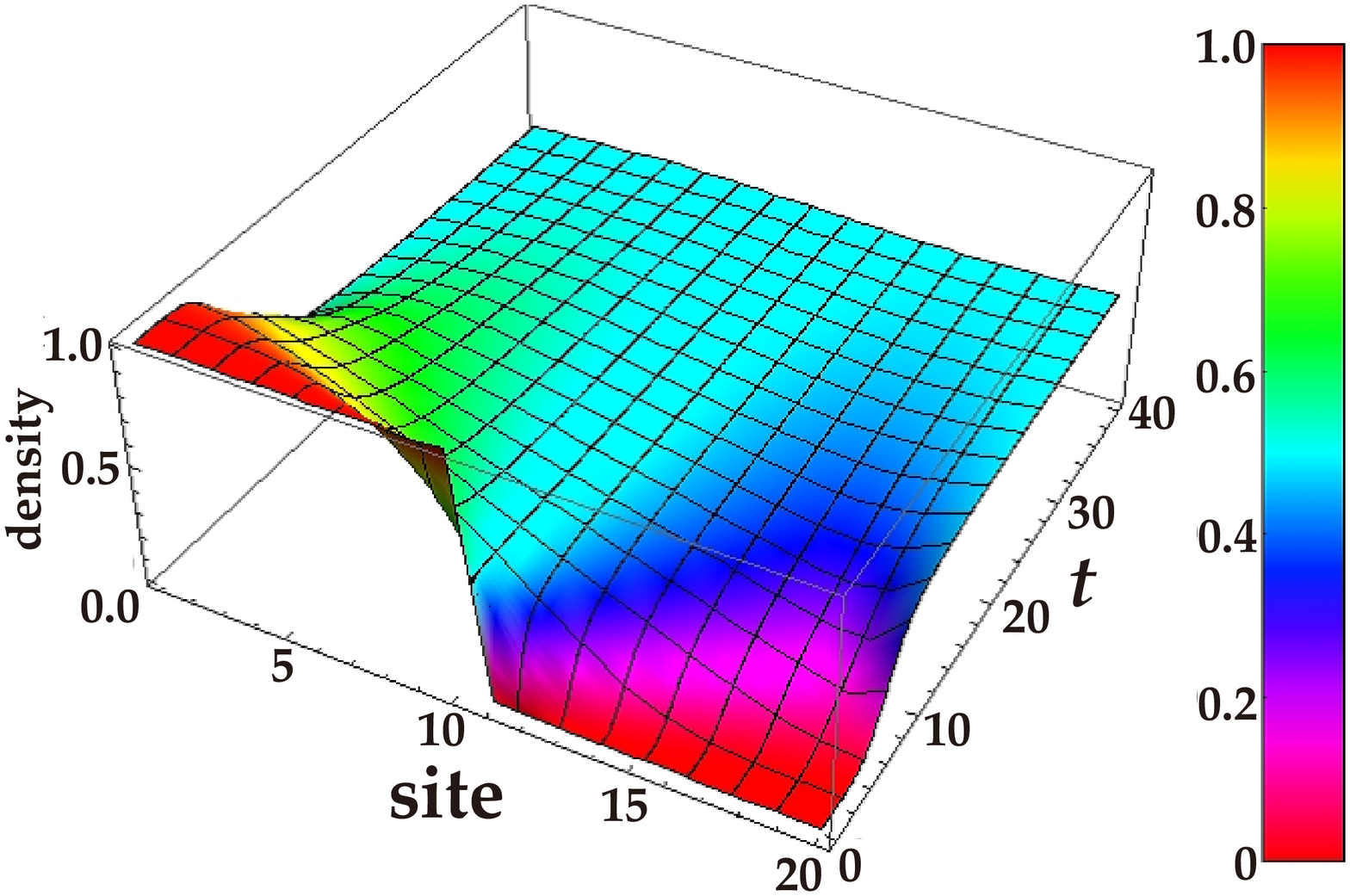}
\includegraphics[width=0.4\textwidth]{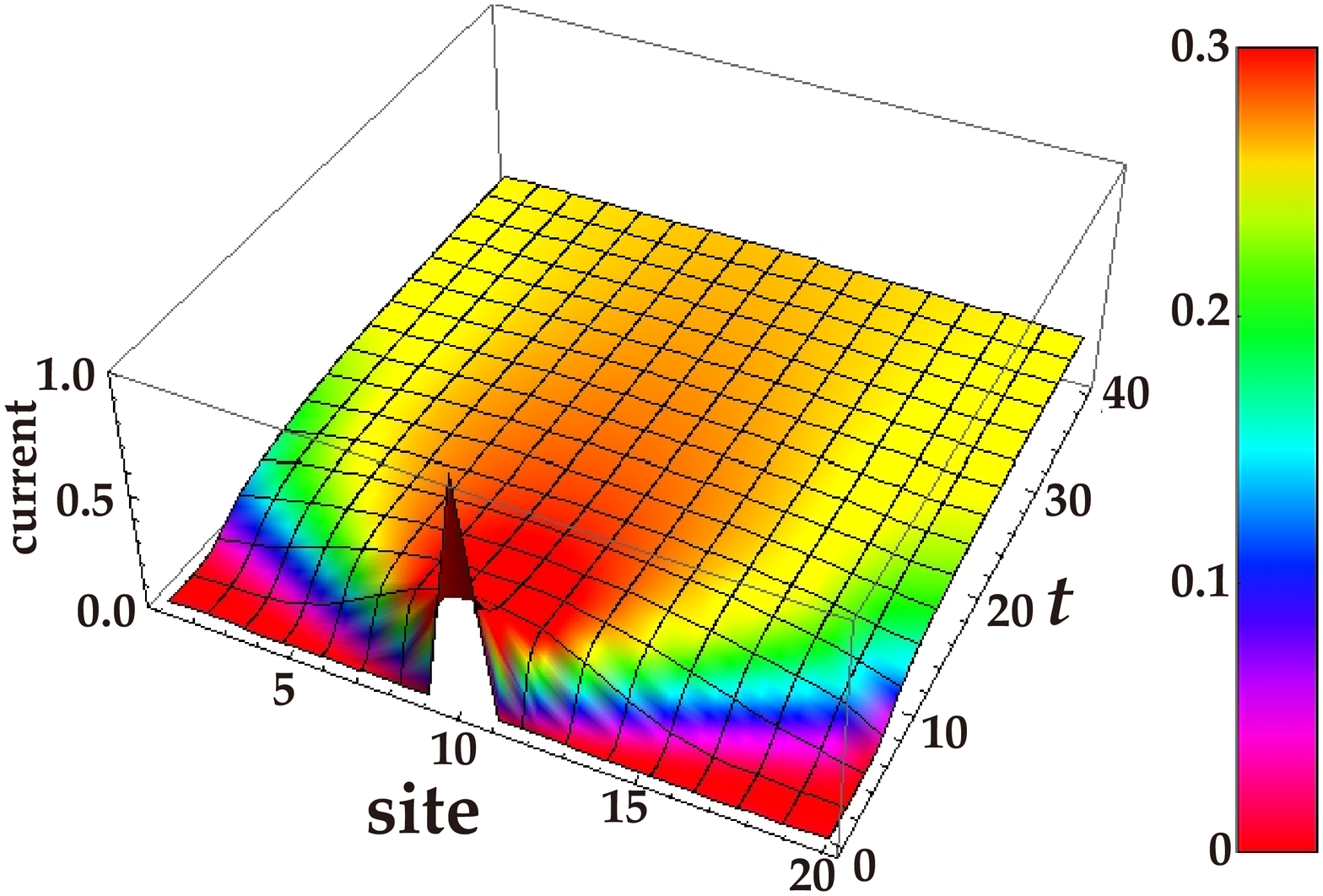}
\end{center}
\caption{(Color) The time evolution of density profile
$\langle n_i \rangle_t$ and current profile
$\langle j_i \rangle_t$ for 20 sites, 10 particles.}
\label{timeevolutiontwo}
\end{figure}

The top panels of Figs. \ref{timeevolution} and \ref{timeevolutiontwo}
show the time evolution of the expectation values of the local particle densities.
In the beginning, the behavior of the local densities matches with our intuition:
as the site is closer to the 10th site where the top particle of the line  occupies,
the local density of a site converges faster to 1/2 which is the density of the 
steady state. This is because the particles begin to flow around the 10th site.
However, some time later, the density on the 1st site becomes to decay 
faster than the 4th site, for example.
This behavior can be intuitively explained:
No particle comes into the 1st site
until the top particle which initially occupies the 10th site
reaches the 20th site.  More precisely, the decay rate
of the expectation value of the particle densities is described
by the continuity equation:
\begin{equation}
\frac{d}{dt} \langle n_k \rangle_t=
\langle j_{k-1} \rangle_t - \langle j_{k} \rangle_t.
\end{equation}
For $k$ is small (say $k=1$), the current does not
flow for a while. After some time, $\langle j_1 \rangle$ becomes visible, but $
\langle j_{20}\rangle (=\langle j_0 \rangle)$ 
is still very small: $\langle j_{20} \rangle 
 \ll \langle j_1 \rangle $ (see the bottom panel of the Fig.~\ref{timeevolution}).
Thus, the density of the 1st site $\langle n_1 \rangle$ decreases in a rapid way
after when the current $\langle j_1\rangle$ begins to flow.
In contrast,  for the sites where the middle particles initially occupy (say $k=4$),
the current $\langle j_3 \rangle$ begins to flow shortly after 
$\langle j_4 \rangle$ has started to flow.
The existence of the entering flow $\langle j_3 \rangle$ 
moderates the decrease of the density 
of the 4th site $\langle n_4 \rangle$. After some time, its value becomes larger than that of 
the 1st site where no particle enters until the top particle reaches.
In this way, we observe the emergence of a ripple in the density profile.

The bottom panels of Figs. \ref{timeevolution} and \ref{timeevolutiontwo}
are the relaxation process of the local currents.
First, note that the current profile is symmetric with respect to the 10th site.
This reflects the fact that we are considering the half-filling case,
and a system with particles moving from left to right with initially occupying $1,2, \cdots, 10$th sites 
can be regarded as a system with holes moving from right to left with initially occupying $11,12, \cdots, 20$th sites.

The current of the 10th site is 1 at $t=0$ since the 11th site is vacant, and monotonically decreases to the steady state current.
The currents around the 10th site show a rapid increase from 0
to reach peak, and then decrease.
These mean that one observes the
excessive flow of currents around the 
10th site which the top particle occupies in the initial state.
As the site is more distant from the 10th site, the increase slope of the current gets flatter, and finally the peak
disappears, i.e., the current no longer flows excessively and
just shows monotonic increase.
The behavior that the currents have peak in some regime between the
1st and 10th sites can be regarded as the combined 
effect of the monotonic decrease represented by the 10th site where the top
particle immediately starts to move
and the monotonic increase represented by
the 1st site where the last particle has to wait
for a while to start moving.

\begin{figure}[tt]
\begin{center}
\includegraphics[width=0.40\textwidth]{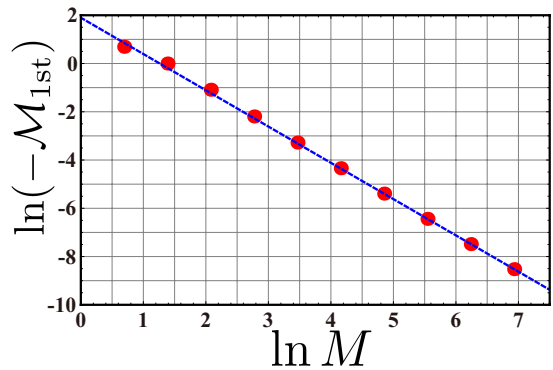}
\end{center}
\caption{(Color online) The lowest excited energy $\mathcal{M}_{\mathrm{1st}}$
plotted against the total number of sites $M$ for $M=2^k,k=1,\cdots,10$.}
\label{scalingenergy}
\end{figure}

{\it Finite-size scaling}.$-$
We conduct the finite-size scaling analysis at half-filling.
For large times, local densities and currents behave as
\begin{align}
\langle n_k \rangle_t& \xrightarrow[t \to \infty]{}
\frac{N}{M}
+
A(n_k) \mathrm{e}^{\mathcal{M}_{\mathrm{1st}}t}, \\
\langle j_k \rangle_t& \xrightarrow[t \to \infty]{}
\frac{N(M-N)}{M(M-1)}
+
A(j_k) \mathrm{e}^{\mathcal{M}_{\mathrm{1st}}t},
\end{align}
where $\mathcal{M}_{\mathrm{1st}}$ is the non-zero eigenvalue
of the Markov matrix $\mathcal{M}$ with largest real part.
First, as a check, we perform the scaling analysis of 
$\mathcal{M}_{\mathrm{1st}}$.
Fig. \ref{scalingenergy} shows $\mathrm{ln}(-\mathcal{M}_{\mathrm{1st}})
\ \mathrm{vs} \ \mathrm{ln}M $.
The simplest fitting from $M=256,512,1024$ gives
$\mathrm{ln}(-\mathcal{M}_{\mathrm{1st}})=1.89793-1.50351 \mathrm{ln}M$,
which implies the KPZ scaling
(see \cite{deGE0,deGE1} for the thorough analysis of the
open as well as the periodic boundary conditions of ASEP).

\begin{figure}[tt]
\begin{center}
\includegraphics[width=0.39\textwidth]{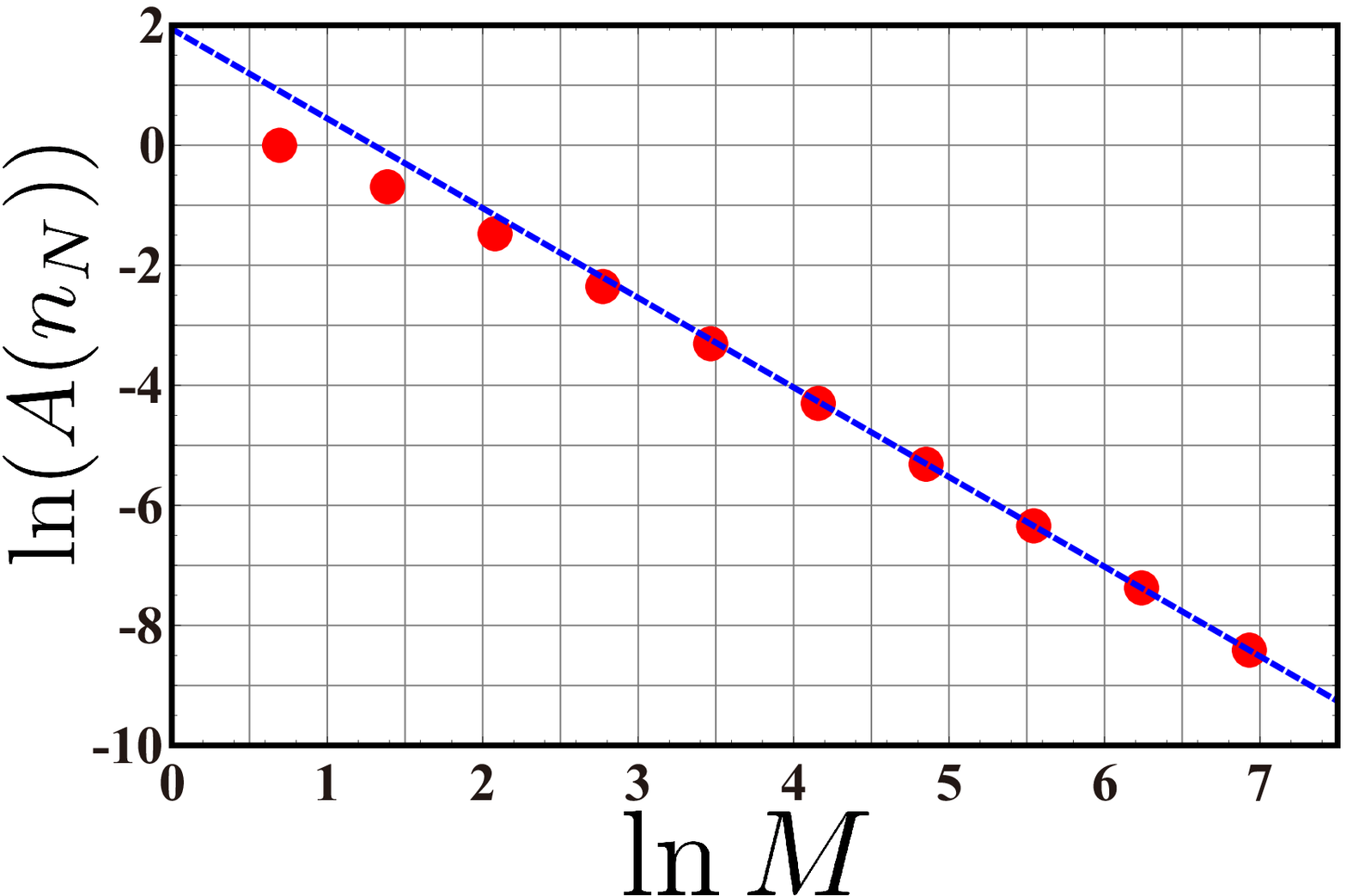}
\includegraphics[width=0.39\textwidth]{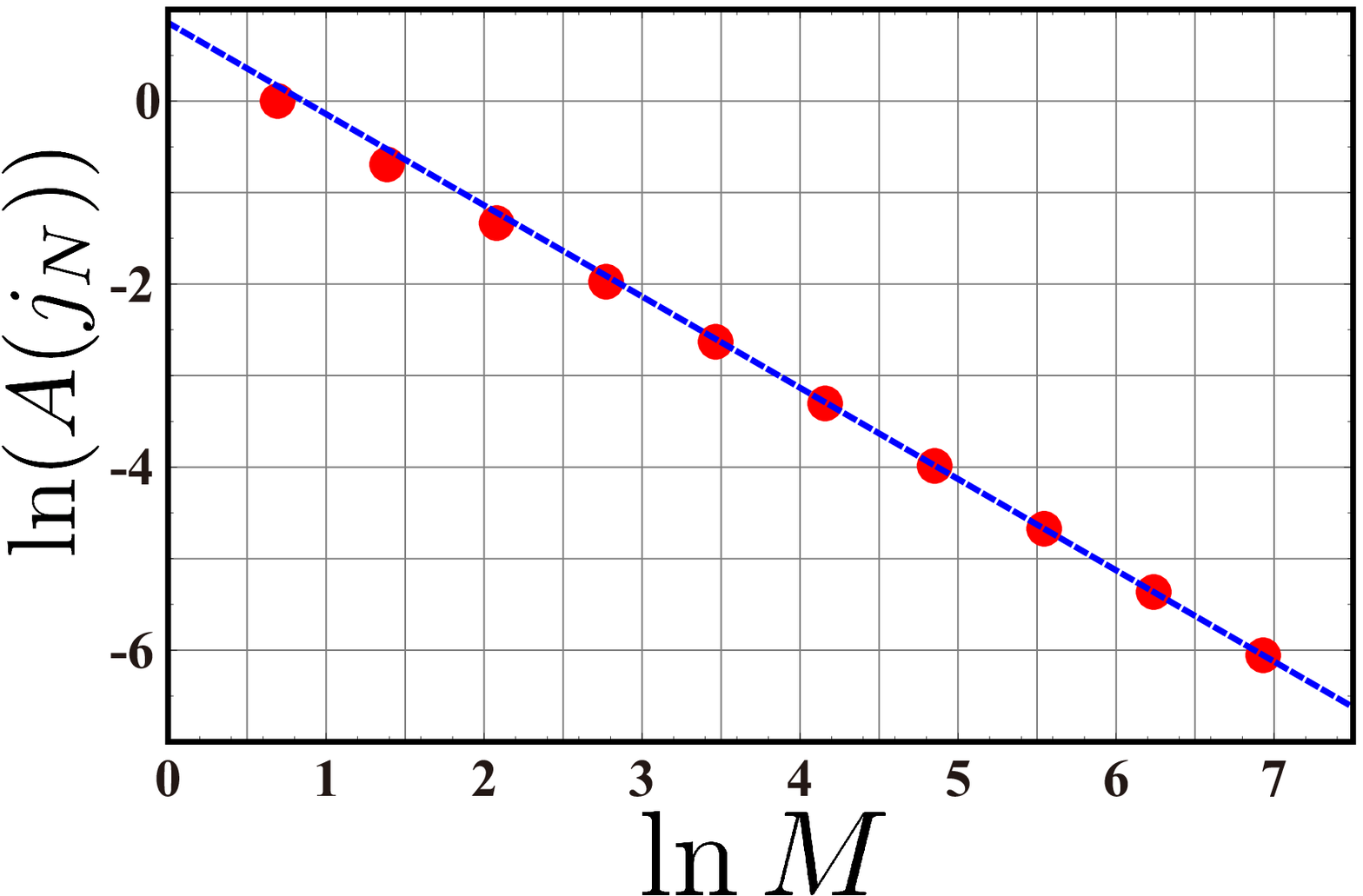}
\end{center}
\caption{(Color online) The asymptotic amplitudes
$A(n_N)$ and $A(j_N)$
of the local density $\langle n_N \rangle_t$
and current $\langle j_N \rangle_t$
plotted against the total number of sites $M$
for $M=2^k,k=1,\cdots,10$.}
\label{dencurrscaling}
\end{figure}

Next, we make the scaling analysis
of the asymptotic amplitudes $A(n_k)$ and $A(j_k)$ of the
local densities $\langle n_k \rangle_t$
and currents $\langle j_k \rangle_t$.
We take the amplitudes at the $N$-th site
$A(n_N)$ and $A(j_N)$ as a representative.
The top and bottom panel of Fig. \ref{dencurrscaling}
shows $\mathrm{ln}(A(n_N)) \ \mathrm{vs} \ \mathrm{ln}M$
and $\mathrm{ln}(A(j_N)) \ \mathrm{vs} \ \mathrm{ln}M$, respectively.
The fitting from $M=256,512,1024$ gives
$\mathrm{ln}(A(n_N))=1.937981-1.4933187 \mathrm{ln}M$
and
$\mathrm{ln}(A(j_N))=0.854409-0.9968579 \mathrm{ln}M$.
Confining to larger sites for fitting,
the coefficient associated with $\mathrm{ln}M$
approaches to $-3/2$ and $-1$, which
indicates $A(n_N) \propto M^{-3/2}$ and $A(j_N) \propto M^{-1}$
respectively.

{\it Conclusions.}$-$
In this article, we have studied the
exact relaxation dynamics of the TASEP.
We examined the behavior of the local densities
and currents by use of the algebraic Bethe ansatz method.
By quantitive evaluation,
we see the convergence of the density and current profiles
to the steady state densities and currents.
Furthermore, we observe interesting behaviors
such as the emergence of a ripple in the density profile
and excessive flow of currents peculiar to the step initial condition.
Moreover, by making a finite-size scaling analysis,
we determine the scaling exponents of the asymptotic amplitudes of 
the local densities and currents with respect to the total number of sites,
which are found to be $-3/2$ and $-1$ respectively.
%
%

The authors would like to thank C. Arita for useful discussions.
The present research
is partially supported by Grant-in-Aid for Young Scientists (B)
No. 21740285. J.S. is supported by JSPS.

\end{document}